\documentclass[12pt]{article}
\usepackage{amsmath}
\usepackage{amssymb}
\usepackage{amsthm}
\usepackage{color}
\allowdisplaybreaks
\newif\ifpersonalnote \personalnotetrue

\newcommand{\personalnoteON}{%
  \ifpersonalnote\personalnotetrue}

\personalnoteON

\title{Discrete Hamiltonians of discrete Painlev\'e equations}

\author{
Takafumi MASE,
Akane NAKAMURA,
and Hidetaka SAKAI
}

\date{}

\begin{document}

\maketitle
\newtheorem{thm}{Theorem}[section]
\newtheorem{prop}[thm]{Proposition}
\def\pf{\noindent{\it Proof.\quad}}
\def\qed{\hfil\fbox{}\medskip}
\theoremstyle{remark}
\newtheorem{rem}{\bf {\slshape Remark}}[section]
\numberwithin{equation}{section}

\begin{abstract}
We express discrete Painlev\'e equations as discrete Hamiltonian systems.
The discrete Hamiltonian systems here mean the canonical transformations defined by generating functions.
Our construction relies on the classification of the discrete Painlev\'e equations based on the surface-type.
The discrete Hamiltonians we obtain are written in the logarithm and dilogarithm functions.


\end{abstract}

\section{Introduction}

At the beginning of the 20th century, P. Painlev\'{e} and B. Gambier classified second order ordinary differential equations of normal form that possess the so-called Painlev\'{e} property \cite{painleve, gambier}.
They discovered six new transcendental equations, which are known today as the Painlev\'{e} equations.
About 80 years later, singularity confinement has been proposed as a discrete analogue of the Painlev\'{e} property \cite{sc} and, with the help of this test, discrete analogues of the Painlev\'{e} equations have been discovered \cite{dpainleve}.
Today, a large number of discrete Painlev\'{e} equations are known, but most of them have been constructed by deautonomizing QRT mappings \cite{hunting,elliptic}.
Since QRT mappings can be solved by elliptic functions \cite{qrt}, this deautonomization procedure is parallel to that in the continuous case, where the (continuous) Painlev\'{e} equations can be thought of as deautonomized systems of ordinary differential equations of elliptic functions.
Using a specific type of rational surfaces, one of the present authors classified (and in a sense, defined) discrete Painlev\'{e} equations \cite{sakai}.
According to this classification, the discrete Painlev\'{e} equations consist of 19 classes depending on the surface type, which we will see later.
The surface associated to an equation is called the space of initial conditions and, through the theory of spaces of initial conditions, both discrete and continuous Painlev\'{e} equations, including their relations, are well-studied.

These days, research on discrete Painlev\'e equations is performed almost in parallel with research on the continuous Painlev\'e equations, such as the reduction to the compatibility conditions of linear equations (Lax pair), the calculation of special solutions, and so on.
One of the biggest difference is that, while the Painlev\'e differential equations are all expressed as Hamiltonian systems, such a description in the discrete case was not yet known.
Let us take a look at the Hamiltonian functions of the Painlev\'e differential equations:
\begin{align}
 & H_\text{VI}\left({a_1, a_2\atop a_3, a_4
 };t;q,p\right)=\;q(q-1)(q-s)p^2\nonumber\\
 &\hspace*{4em}\!\!\! +\Big\{ (a_1\! +\! 2a_2) q(q\! -\! 1)\!
 +\! a_3(s\! -\! 1)q\! +\! a_4s(q\! -\! 1)\Big\} p
+a_2(a_1\! +\! a_2) q,\qquad
\frac{ds}{dt}=s(s-1),
 \nonumber\\
& H_\text{V}\left({a_1, a_2\atop a_3};t;q,p\right)
=\;p(p+e^t)q(q-1)
-a_1p(q-1)-a_3 pq+a_2e^tq,\nonumber\\
& H_\text{III}(D_6)\left(a_1, b_1;t;q,p\right)=\; 
p(p+1)q^2-a_1p(q-1)-b_1pq-tq,
\nonumber\\ 
& H_\text{III}(D_7)\left(a_1;t;q,p\right)=\; 
p^2q^2+a_1qp+e^tp+q,
\qquad
H_\text{III}(D_8)\left(t;q,p\right)=
p^2q^2+qp-q-\frac{e^t}{q},
\nonumber\\
&H_\text{IV}\left(a_1, a_2;t;q,p\right)=
pq(p-q-t)-a_1 p-a_2 q,\nonumber\\
&\label{eqn:pain_ham}
 H_\text{II}\left(a_1;t;q,p\right)=\; 
p(p-q^2-t)-a_1 q,\qquad\quad
 H_\text{I}\left(t;q,p\right)=\; 
p^2-q^3-tq.
\end{align}

Expressing an equation as a Hamiltonian system has many advantages.
One of the most important benefit is that the Hamiltonian function automatically becomes a conserved quantity if it is autonomous.
In the case of the Painlev\'{e} equations, however, this does not hold since the systems are non-autonomous.
Another important advantage is that using a Hamiltonian system we can write an equation concisely.
For example, this can be of help when one considers the problem of identifying equations.
Since the time evolution is determined by a single function, one can compare the Hamiltonian functions instead of the time evolution equations themselves.

Roughly speaking, when a discrete dynamical system is ``easily'' expressed by a single function $W$ on some phase space, we call $W$ a ``discrete Hamiltonian'' of the system.
As an example of such a function, we already know what is called the generating function of a canonical transformation.

A canonical transformation on a phase space with a symplectic structure is defined as a transformation of the Hamiltonian system that preserves the symplectic form.
It is known that each canonical transformation can be written with a function $W=W(q, \overline{p})$ as
\begin{align}
 &p_k=\frac{\partial W}{\partial q_k},\qquad
 \overline{q}_k=\frac{\partial W}{\partial \overline{p}_k},\qquad k=1,\ldots, n,
\end{align}
where $W$ is called the generating function.
In the case of discrete Painlev\'{e} equations, however, it is usually more important to write a system as a birational mapping than to write in canonical variables.
Therefore, in this paper, we sometimes give priority to choosing good variables over writing discrete Hamiltonians or equations in canonical variables

\begin{rem}
Let us take a look at a relation with the Lagrangian form of discrete dynamical systems by Veselov \cite{veselov}.
 Given a Legendre function $L_k (r,s):\ X\times X\rightarrow \mathbb{C}$, the variation of the formal sum $S(\lambda)=\sum_{k\in\mathbb{Z}}L_k (\lambda_k,\lambda_{k+1})$
 \begin{align*}
 \delta S(\lambda)=&\sum_{k\in\mathbb{Z}}\delta L_k (\lambda_k,\lambda_{k+1})\\
 =&\sum_{k\in\mathbb{Z}}\left\{ L_k (\lambda_k+\delta \lambda_k,\lambda_{k+1}+\delta \lambda_{k+1})
 -L_k (\lambda_k,\lambda_{k+1})\right\}\\
 =&\sum_{k\in\mathbb{Z}}\left\{
 \frac{\partial L_k}{\partial r}(\lambda_k,\lambda_{k+1})\delta \lambda_k
 +\frac{\partial L_k}{\partial s}(\lambda_k,\lambda_{k+1})\delta \lambda_{k+1}\right\}\\
 =&\sum_{k\in\mathbb{Z}}\left\{
 \frac{\partial L_k}{\partial r}(\lambda_k,\lambda_{k+1})
 +\frac{\partial L_{k-1}}{\partial s}(\lambda_{k-1},\lambda_k)\right\} \delta \lambda_k =0
\end{align*}
 gives the discrete Euler-Lagrange equation
 \begin{align}
  &\frac{\partial L_k}{\partial r}(\lambda_k,\lambda_{k+1})
  +\frac{\partial L_{k-1}}{\partial s}(\lambda_{k-1},\lambda_k) =0.
 \end{align}
This equation is a second-order single equation.
Let us rewrite it into a simultaneous form.
Putting $\mu_k=\frac{\partial L_k}{\partial r}(\lambda_k,\lambda_{k+1})=
 -\frac{\partial L_{k-1}}{\partial s}(\lambda_{k-1},\lambda_k)$
and $W(\lambda,\overline{\mu})=\overline{\lambda}\overline{\mu}+L(\lambda,\overline{\lambda})$,
we can write the equation as
 \begin{align}
  &  \overline{\lambda}=\frac{\partial W}{\partial\overline{\mu}},\quad
  \mu =\frac{\partial W}{\partial\lambda},
 \end{align}
which is an expression as a canonical transformation by the generating function $W$.
In fact, the symplectic form $d\mu\wedge d\lambda$ is preserved under this transformation.

In this paper, we do not consider the Lagrangian form but focus on generating functions of canonical transformations.
\qed
\end{rem}

It is known that each discrete Painlev\'e equation can be formulated as a discrete dynamical system determined by a Cremona isometry of infinite order on a generalized Halphen surface \cite{sakai}.
Generalized Halphen surfaces are classified according to the type of the anti-canonical divisor.
The list of the surfaces is as in Table~\ref{table:ghs}.
\begin{table}[h]\label{table:ghs}
\begin{center}
 \begin{tabular}{|c|c|c|} \hline
 elliptic & multiplicative & additive \\ \hline
 $A_0^{(1)}$ & $A_0^{(1)\ast}\!\!$,\, $A_1^{(1)}\!\!$,\, $A_2^{(1)}\!\!$,\,
 & $A_0^{(1)\ast\ast}\!\!$,\, $A_1^{(1)\ast}\!\!$,\,
 $A_2^{(1)\ast}\!\!$,\, \\
 & $A_3^{(1)}\!\! ,\,\ldots , A_6^{(1)}\!\!$,\, $A_7^{(1)}\!\!$,\,
 $A_7^{(1)\prime}\!\!$,\, $A_8^{(1)}$
 & $D_4^{(1)}\!\!$,\, $D_5^{(1)}\!\!$,\, $D_6^{(1)}\!\!$,\,
 $D_7^{(1)}\!\!$,\, $D_8^{(1)}\!\!$,\,\\
 && $E_6^{(1)}\!\!$,\, $E_7^{(1)}\!\!$,\, $E_8^{(1)}$ \\ \hline
 \end{tabular}
 \caption{List of generalized Halphen surfaces}
 \end{center}
 \end{table}
  \vspace{-5mm}
 \\
The surfaces of type $A_8^{(1)}$, $D_8^{(1)}$, and $E_8^{(1)}$ have no Cremona isometries of infinite order.
That is, there are no discrete Painlev\'e equations in these cases.
In addition, all the surfaces other than of type $E_8^{(1)}$ have a blowing-down to $\mathbb{P}^1\times\mathbb{P}^1$.
In most cases, the image of the anti-canonical divisor can be taken as ${f_0}^2{g_0}^2=0$,
$f_0f_1g_0g_1=0$, or $f_0f_1{g_0}^2=0$ on $\mathbb{P}^1 \times \mathbb{P}^1$,
where $ (f_0: f_1), (g_0: g_1) $ are a bi-homogeneous coordinate.
The exceptions are of type
$A_2^{(1)\ast}$,
$A_2^{(1)}$,
$A_1^{(1)\ast}$,
$A_1^{(1)}$,
$A_0^{(1)\ast\ast}$,
$A_0^{(1)\ast}$,
and $A_0^{(1)}$.

In \textsection\ref{sec:biquadratic}--\ref{sec:ushape}, we consider the three regular cases, respectively.
We first look at concrete forms of discrete systems and then we write them as discrete Hamiltonian systems.
In \textsection\ref{sec:exceptional}, we consider the exceptional cases.
We will only look at a specific calculation of type $A_2^{(1)}$.
It should be noted, however, that the discrete Painlev\'e equations we will see here are nothing but well-known representatives for each surface and that each surface can have an infinite number of different discrete equations.

\section{The case: ${f_0}^2{g_0}^2=0$}\label{sec:biquadratic}
%
%
These are cases where the image of the anti-canonical divisor can be chosen as ${f_0}^2{g_0}^2=0$.
The surfaces of type $D_5^{(1)}$, $D_6^{(1)}$, $D_7^{(1)}$, $E_6^{(1)}$, and $E_7^{(1)}$ fall into this category.
In addition to discrete equations, the differential Painlev\'e equations arise from these surfaces.
Using the inhomogeneous coordinate $f = f_1/f_0$ and $g = g_1/g_0$, the
Hamiltonians of these differential Painlev\'e equations are expressible in biquadratic forms:
  \begin{align}
  &\frac{df}{dt}=\frac{\partial H}{\partial g},\quad
  \frac{dg}{dt}=-\frac{\partial H}{\partial f},\nonumber\\
   &H=(g^2, g, 1)\left(\begin{array}{ccc}
       m_{22} & m_{21} & m_{20} \\
       m_{12} & m_{11} & m_{10} \\
       m_{02} & m_{01} & m_{00}
	    \end{array}\right)\left(\begin{array}{c}
	    f^2\\
	    f\\
            1
				    \end{array}\right),
  \end{align}
%
where the matrix $M=(m_{ij})_{i,j=2,1,0}$ can be chosen as follows:
\begin{align}
 M=&M_{D_5}=\left(\begin{array}{ccc}
	    1 & -1 & 0 \\
	    s & -a_1-a_3-s & a_1 \\
	    0 & sa_2 & 0
	       \end{array}\right) ,\quad
 M_{D_6}=\left(\begin{array}{ccc}
	    1 & 0 & 0 \\
	    1 & -a_1-b_1 & -a_1 \\
	    0 & -s & 0
		  \end{array}\right),\nonumber\\
 &\!\!\!\! M_{D_7}=\!\left(\!\begin{array}{ccc}
	    1 & 0 & 0 \\
	    0 & a_1 &\!\! -1\! \\
	    0 & s & 0
		 \end{array}\!\right)\! ,\quad\!\!
 M_{E_6}=\!\left(\!\begin{array}{ccc}
	    0 & 1 & 0 \\
	    \!\! -1\! & \!\! -s\! & \!\! -a_2\! \\
	    0 & \!\! -a_1\! & 0
		 \end{array}\!\right)\! ,\quad\!\!
 M_{E_7}=\!\left(\!\begin{array}{ccc}
	    0 & 0 & 1 \\
	    \!\! -1\! & 0 & \!\! -s\! \\
	    0 & \!\! -a_1\! & 0
		 \end{array}\!\right)\! .
\end{align}
%
Discrete Painlev\'e systems can be expressed in terms of these matrices $M=(m_{ij})_{i,j=2,1,0}$ as
\begin{align}\label{eqn:biquad}
 &g=-\overline{g}-\frac{m_{12}f^2+m_{11}f+m_{10}}{m_{22}f^2+m_{21}f+m_{20}},\qquad
 \overline{f}=-f-\frac{\overline{m}_{21}{\overline{g}}^2+\overline{m}_{11}\overline{g}+\overline{m}_{01}}
 {\overline{m}_{22}{\overline{g}}^2+\overline{m}_{12}\overline{g}+\overline{m}_{02}}.
\end{align}
%
Defining the generating function $W$ by
\begin{align}
 &W=W(f,\overline{g})=-f\overline{g}-\int\frac{m_{12}f^2+m_{11}f+m_{10}}{m_{22}f^2+m_{21}f+m_{20}}df
 -\int\frac{\overline{m}_{21}{\overline{g}}^2+\overline{m}_{11}\overline{g}+\overline{m}_{01}}
 {\overline{m}_{22}{\overline{g}}^2+\overline{m}_{12}\overline{g}+\overline{m}_{02}}d\overline{g},
\end{align}
the discrete system (\ref{eqn:biquad}) is expressed as
\begin{align}
 &g=\frac{\partial W}{\partial f},\qquad
 \overline{f}=\frac{\partial W}{\partial\overline{g}}.
\end{align}

The explicit forms of these discrete Hamiltonians for each type are
\begin{align}
	W =
 	&W_{\! D_5} = 
	- f \overline{g} - f s + \overline{g} + a_3 \log(f - 1) + a_1 \log f - \overline{a}_2 \log \overline{g} + (\overline{a}_1 + \overline{a}_2 + \overline{a}_3) \log (\overline{g} + s), \\
 	&W_{\! D_6} = 
	- f \overline{g} - f - \frac{a_1}{f} + (a_1 + b_1) \log f + s \log \overline{g} + (\overline{a}_1 + \overline{b}_1 - s) \log (\overline{g} + 1), \\
	&W_{\! D_7} = 
	- f \overline{g} - \frac{1}{f} + \frac{s}{\overline{g}} - a_1 \log f - \overline{a}_1 \log \overline{g}, \\
	&W_{\! E_6} = 
	- f \overline{g} + \frac{f^2}{2} + s f + \frac{\overline{g}^2}{2} - s \overline{g} + a_2 \log f - \overline{a}_1 \log \overline{g}, \\
	&W_{\! E_7} = 
	- f \overline{g} + s f + \frac{f^3}{3} - \overline{a}_1 \log \overline{g}.
\end{align}
For instance, when the surface is of type $E_7^{(1)}$, the discrete Hamiltonian $W_{\! E_7}$ gives the system
 \begin{align*}
  &g=\frac{\partial W_{\! E_7}}{\partial f}=-\overline{g}+s+f^2,\\
 &\overline{f}=\frac{\partial W_{\! E_7}}{\partial\overline{g}}=-f-\frac{\overline{a}_1}{\overline{g}},
 \end{align*}
which is in fact a discrete Painlev\'e system of type $E_7^{(1)}$.

\section{The case: $f_0f_1g_0g_1=0$}
%
These are cases when the image of the anti-canonical curve can be chosen as $f_0f_1g_0g_1=0$.
The surfaces of type $A_3^{(1)}$, $A_4^{(1)}$, $A_5^{(1)}$, $A_6^{(1)}$, $A_7^{(1)}$, and $A_7^{(1)\prime}$ fall into this category.
There are no differential Painlev\'e equations attached to these surfaces.
Discrete Painlev\'e systems are expressed with a $3$ by $3$ matrix $M=(m_{ij})_{i,j=2,1,0}$ as
\begin{align}\label{eqn:i_gata}
 &g=\frac{m_{02}f^2+m_{01}f+m_{00}}{\overline{g}(m_{22}f^2+m_{21}f+m_{20})},\qquad
 \overline{f}=\frac{\overline{m}_{20}{\overline{g}}^2+\overline{m}_{10}\overline{g}+\overline{m}_{00}}
 {f(\overline{m}_{22}{\overline{g}}^2+\overline{m}_{12}\overline{g}+\overline{m}_{02})}.
\end{align}
The symplectic form is $\omega =\frac{dg\wedge df}{fg}=d\log g\wedge d\log f$.
Using $G=\log g$ and $F=\log f$, the system can be expressed as
\begin{align}
 &G=-\overline{g}+\log\left(m_{02}e^{2F}+m_{01}e^F+m_{00}\right)
 -\log\left(m_{22}e^{2F}+m_{21}e^F+m_{20}\right),\nonumber\\
 &\overline{f}=-F+\log\left(\overline{m}_{20}e^{2\overline{g}}
 +\overline{m}_{10}e^{\overline{g}}+\overline{m}_{00}\right)
 -\log\left(\overline{m}_{22}e^{2\overline{g}}+\overline{m}_{12}e^{\overline{g}}+\overline{m}_{02}\right).
\end{align}
%
Defining the generating function $\widetilde{W}$ by
\begin{align}
 \widetilde{W}=\widetilde{W}(F,\overline{g})=&-F\overline{g}
 +\int\log\left(m_{02}e^{2F}+m_{01}e^F+m_{00}\right)dF
 -\int\log\left(m_{22}e^{2F}+m_{21}e^F+m_{20}\right)dF\nonumber\\
 &+\int\log\left(\overline{m}_{20}e^{2\overline{g}}+\overline{m}_{10}e^{\overline{g}}+\overline{m}_{00}\right)d\overline{g}
 -\int\log\left(\overline{m}_{22}e^{2\overline{g}}+\overline{m}_{12}e^{\overline{g}}+\overline{m}_{02}\right)d\overline{g},
\end{align}
the discrete system (\ref{eqn:biquad}) is written as
\begin{align}
 &G=\frac{\partial\widetilde{W}}{\partial F},\qquad
 \overline{f}=\frac{\partial\widetilde{W}}{\partial\overline{g}}.
\end{align}
In order to write the system in $f$ and $g$, we introduce $W(f,\overline{g})=\widetilde{W}(\log f,\log\overline{g})$:
\begin{align}
 W(f,\overline{g})=&-\log f\log\overline{g}
 +\int\log\left(m_{02}f^2+m_{01}f+m_{00}\right)\frac{df}{f}
 -\int\log\left(m_{22}f^2+m_{21}f+m_{20}\right)\frac{df}{f}\nonumber\\
 &+\int\log\left(\overline{m}_{20}{\overline{g}}^2+\overline{m}_{10}\overline{g}+\overline{m}_{00}\right)\frac{d\overline{g}}{\overline{g}}
 -\int\log\left(\overline{m}_{22}{\overline{g}}^2+\overline{m}_{12}\overline{g}+\overline{m}_{02}\right)\frac{d\overline{g}}{\overline{g}},
\end{align}
with which the discrete system (\ref{eqn:i_gata}) is expressed as
\begin{align}
 &g=\exp\left(f\frac{\partial W}{\partial f}\right),\qquad
 \overline{f}=\exp\left(\overline{g}\frac{\partial W}{\partial\overline{g}}\right).
\end{align}

For each type of surface, the matrices $M$ can be chosen as
\begin{align}
 &
 M_{A_7}=\left(\begin{array}{ccc}
	    0 & -a_0 & 0 \\
	    1 & 0 & 0 \\
	    0 & -1 & 1
	     \end{array}\right) ,\quad
M_{A_7^\prime}=\left(\begin{array}{ccc}
	    1 & -a_0 & 0 \\
	    0 & 0 & 0 \\
	    0 & -1 & 1
		 \end{array}\right) ,\nonumber\\
&M_{A_6}=\left(\begin{array}{ccc}
	    0 & 1/b & 0 \\
	    1 & 0 & -1/b \\
	    0 & -a_1 & a_1
		 \end{array}\right) ,\quad
 M_{A_5}=\left(\begin{array}{ccc}
	    0 & b_1/a_2 & 0 \\
	    a_0 & 0 & -b_1/a_2 \\
	    1/a_1 & -1-(1/a_1) & 1
		 \end{array}\right) ,\nonumber\\
&M_{A_4}=\left(\begin{array}{ccc}
	    0 & 1 & -1 \\
	    a_0/a_2 & 0 & 1+(1/a_4) \\
	    -a_0a_3/a_2 & a_0a_3+(1/a_2a_4) & -1/a_4
		 \end{array}\right) ,\nonumber\\
&M_{A_3}=\left(\begin{array}{ccc}
	    a_0a_5 & -(1+a_0)a_5 & a_5 \\
 -1/a_1{a_2}^2a_3-a_0a_3a_5& 0 & -1-a_5 \\
	    1/(a_1{a_2}^2) & -(1+a_1)/a_1a_2 & 1
		 \end{array}\right).
\end{align}
The corresponding discrete Hamiltonians can be explicitly written as
\begin{align}
	W =
	&W_{\! A_3} = 
 -\log f\log\overline{g}
 +\mathrm{Li}_2(f)
 +\mathrm{Li}_2(a_0f)
 -\mathrm{Li}_2\left(\frac{f}{a_2}\right)
 -\mathrm{Li}_2\left(\frac{f}{a_1a_2}\right)
 \nonumber
 \\
 &\hspace{10mm}
 -\mathrm{Li}_2\overline{g}
 +\mathrm{Li}_2\left(\frac{\overline{g}}{\overline{a}_3}\right)
 -\mathrm{Li}_2\left(\overline{a}_5\overline{g}\right)
 +\mathrm{Li}_2\left(\overline{a}_0\overline{a}_1\overline{a}_2^2\overline{a}_3\overline{a}_5\overline{g}\right)
 -\log a_5\log f +\log(\overline{a}_1{\overline{a}_2}^2)\log\overline{g},
 \\
	&W_{\! A_4} = 
	-\log{f}\log\overline{g} + \mathrm{Li}_2 (f) - \mathrm{Li}_2 \left( \frac{f}{a_2} \right) - \mathrm{Li}_2 (a_0 a_3 a_4 f) \nonumber\\
	&\hspace{5em} 
 - \mathrm{Li}_2 (\overline{g}) - \mathrm{Li}_2 (\overline{a}_4 \overline{g}) + \mathrm{Li}_2 \left( \frac{\overline{g}}{\overline{a}_3} \right)-\log a_4\log{f} + \left( \log \frac{\overline{a}_2}{\overline{a}_0 \overline{a}_3 \overline{a}_4} \right)\log\overline{g}, \\
	&W_{\! A_5} = 
	-\log{f}\log\overline{g}- \mathrm{Li}_2 (f) - \mathrm{Li}_2 \left( \frac{f}{a_1} \right)\nonumber\\
        &\hspace{5em} - \mathrm{Li}_2 \left( \frac{\overline{b}_1 \overline{g}}{\overline{a}_2} \right ) + \mathrm{Li}_2 (- \overline{a}_0 \overline{a}_1 \overline{g}) - \frac{1}{2} \left( \log \frac{b_1 f}{a_1} \right)^2+\log\overline{a}_1\log\overline{g}, \\
	&W_{\! A_6} = 
	-\log{f}\log\overline{g} - \mathrm{Li}_2 (f) - \mathrm{Li}_2 \left( \frac{\overline{g}}{\overline{a}_1 \overline{b}} \right)\nonumber\\
        &\hspace{5em} - \frac{1}{2} \left( \log \frac{f}{b} \right)^2 - \frac{1}{2} (\log \overline{g})^2 +\log a_1\log f+ \log\overline{a}_1\log\overline{g}, \\
	&W_{\! A'_7} = 
        -\log{f}\log\overline{g}
   -\mathrm{Li}_2(f)
   +\mathrm{Li}_2\left(\frac{f}{a_0}\right)- \frac12\left(\log f\right)^2 - \left(\log \overline{g}\right)^2
   -\log a_0\log f,\\
	&W_{\! A_7} = 
	- \log f\log \overline{g}- \mathrm{Li}_2 (f) - \frac{1}{2} (\log(- a_0 f))^2 - \frac{1}{2} (\log\overline{g})^2,
\end{align}
where $\mathrm{Li}_2(x)$ is the dilogarithm function 
\begin{align*}
 \mathrm{Li}_2(x)
 =-\int\frac{\log(1-x)}{x}dx
 =\sum_{k=1}^{\infty}\frac{x^k}{k^2}.
\end{align*}
For instance, the discrete Hamiltonian $W_{\! A'_7}$ gives rise to
 \begin{align*}
   g=&\exp\left(f\frac{\partial W_{\! A_7^\prime}}{\partial f}\right)
 =\exp\left(-\log{\overline{g}-\log f+\log
  (1-f)-\log\left(1-\frac{f}{a_0}\right)}-\log a_0\right)
  =\frac{1-f}{\overline{g}f\left(a_0-f\right)},\\
 \overline{f}=&\exp\left(\overline{g}\frac{\partial
  W_{\! A_7^\prime}}{\partial\overline{g}}\right)
  =\exp\left(-\log f-2\log\overline{g}\right)
  =\frac{1}{f\overline{g}^2},
 \end{align*}
which is in fact a discrete Painlev\'e system of type $A_7^{(1)\prime}$.

\section{The case: $f_0f_1{g_0}^2=0$}\label{sec:ushape}

These are cases when the image of the anti-canonical curve can be chosen as $f_0f_1{g_0}^2=0$.
The surfaces of type $D_4^{(1)}$, $D_5^{(1)}$, $D_6^{(1)}$, $D_7^{(1)}$, and $D_8^{(1)}$ fall into this category.
Discrete Painlev\'e systems are expressed with a $3$ by $3$ matrix $M=(m_{ij})_{i,j=2,1,0}$ as
\begin{align}\label{eqn:ko_gata}
 &g=-\overline{g}-\frac{m_{12}f^2+m_{11}f+m_{10}}{m_{22}f^2+m_{21}f+m_{20}},\qquad
 \overline{f}=\frac{\overline{m}_{20}{\overline{g}}^2+\overline{m}_{10}\overline{g}+\overline{m}_{00}}
 {f(\overline{m}_{22}{\overline{g}}^2+\overline{m}_{12}\overline{g}+\overline{m}_{02})}.
\end{align}
Defining the generating function $W$ by
\begin{align}
 W=W(f,\overline{g})=&-\overline{g}\log f-\int\frac{m_{12}f^2+m_{11}f+m_{10}}{m_{22}f^2+m_{21}f+m_{20}}\frac{df}{f}
 \nonumber\\
 &+\int\log\left(\overline{m}_{20}{\overline{g}}^2+\overline{m}_{10}\overline{g}+\overline{m}_{00}\right){d\overline{g}}
 -\int\log\left(\overline{m}_{22}{\overline{g}}^2+\overline{m}_{12}\overline{g}+\overline{m}_{02}\right){d\overline{g}},
\end{align}
the discrete system (\ref{eqn:ko_gata}) is expressed as
\begin{align}
 &g=f\frac{\partial W}{\partial f},\qquad
 \overline{f}=\exp\left(\frac{\partial W}{\partial\overline{g}}\right).
\end{align}

The surface of type $D_8^{(1)}$ does not possess any discrete system.
We have already treated the cases $D_5^{(1)}$, $D_6^{(1)}$, and $D_7^{(1)}$ in the previous section.
Therefore, let us consider the remaining case: type of $D_4^{(1)}$.
The matrix $M$ can be chosen as follows:
\begin{align}
 & M_{D_4}=\left(\begin{array}{ccc}
   1 & -1-s & s \\
   a_1+2a_2 & -a_1-2a_2+(s-1)a_3+sa_4
   & -sa_4 \\
   a_2(a_1+a_2) & 0 & 0
	 \end{array}\right).
\end{align}
Note that although the $D_4^{(1)}$-type surface also possesses a differential equation, the Hamiltonian (\ref{eqn:pain_ham}) is not completely the same as the biquadratic one defined by $M_{D_4}$ since the canonical variables are not $g$ and $\log f$ but $p=g/f$ and $q=f$.
The Hamiltonian in the continuous case is given by $H_\mathrm{VI}=\frac{1}{q}(q^2p^2, qp, 1)M_{D_4}
{\scriptsize
\left(\begin{array}{c}
 q^2 \\
 q \\
 1
\end{array}\right)
}$.

In the case of $D_4^{(1)}$, the discrete Hamiltonian is given by
\begin{align}
 W_{\! D_4}=&-\overline{g}\log f
 +a_4\log f
 -a_3\log (1-f)
 -\left(a_1+2a_2+a_3+a_4\right)\log (1-f/s)
 +\overline{g}(\log\overline{g}+\log s)\nonumber\\
 &-(\overline{g}+\overline{a}_{1}+\overline{a}_{2})\log(\overline{g}+\overline{a}_{1}+\overline{a}_{2})
 -(\overline{g}+\overline{a}_{2})\log(\overline{g}+\overline{a}_{2})
 +(\overline{g}-\overline{a}_{4})\log(\overline{g}-\overline{a}_{4}).
\end{align}
The discrete system determined by $W_{\! D_4}$ is
 \begin{align*}
   g=&f\frac{\partial W_{\! D_4}}{\partial f}
 =-\overline{g}+a_4+\frac{a_3 f}{1-f}+\frac{(a_1+2a_2+a_3+a_4)f}{s-f}\\
 =&-\overline{g}-a_1-2a_2-2a_3+\frac{a_3}{1-f}+\frac{a_1+2a_2+a_3+a_4}{1-f/s}\\
 \overline{f}=&\exp\left(\frac{\partial
  W_{\! D_4}}{\partial\overline{g}}\right)
  =\exp\left(-\log f+\log\overline{g}+\log (\overline{g}-\overline{a}_4) +\log s
  -\log (\overline{g}+\overline{a}_1+\overline{a}_2)
  -\log (\overline{g}+\overline{a}_2)\right)\\
  =&\frac{s\overline{g}(\overline{g}-\overline{a}_4)}
  {f(\overline{g}+\overline{a}_1+\overline{a}_2)(\overline{g}+\overline{a}_2)},
 \end{align*}
which is in fact a discrete Painlev\'e system of type $D_4^{(1)}$.

\section{The other cases}\label{sec:exceptional}

The surfaces of type $A_2^{(1)\ast}$,
$A_2^{(1)}$,
$A_1^{(1)\ast}$,
$A_1^{(1)}$,
$A_0^{(1)\ast\ast}$,
$A_0^{(1)\ast}$, and
$A_0^{(1)}$ fall into this category.
Elliptic difference systems arise for the $A_0^{(1)}$-type surface but, at this moment, a discrete Hamiltonian is difficult to write down for some technical reasons.
Therefore, we do not consider the case $A_0^{(1)}$.

In the other cases, we can take $F=F(f,g)$, $G=G(f,g)$ so that the symplectic form can be expressed as $\omega =d\log G\wedge d\log F$ or $dG\wedge d\log F$.
Discrete Painlev\'{e} systems are expressed by a generating function $W=W(F,\overline{G})$ as
\begin{align}\label{eqn:add_discrete}
 &G=F\frac{\partial W}{\partial F},\qquad
 \overline{F}=\exp\left(\frac{\partial W}{\partial\overline{G}}\right)
\end{align}
or
\begin{align}\label{eqn:mul_discrete}
 &G=\exp\left(F\frac{\partial W}{\partial F}\right),\qquad
 \overline{F}=\exp\left(\overline{G}\frac{\partial W}{\partial\overline{G}}\right).
\end{align}
Such a ${W}$ is an algebraic function in $F$ and $\overline{G}$ but does not have concise expression since it is not single valued.

However, if the system has the form $g=\varphi (f,\overline{g})$, $\overline{f}=\psi (f,\overline{g})$,  $\widetilde{W}$ has a concise expression in $f$, $\overline{g}$.
Here, we use the  notation $\widetilde{W}(f,\overline{g})=W(F(f,\psi (f,\overline{g})),\overline{G}(\varphi(f,\overline{g}),\overline{g}))$.
Nevertheless, even with $\widetilde{W}(f,\overline{g})$, the expressions are not as simple as we want.

While the additive type surfaces $A_2^{(1)\ast}$,
$A_1^{(1)\ast}$, and
$A_0^{(1)\ast\ast}$
correspond to \eqref{eqn:add_discrete},
the multiplicative type surface $A_2^{(1)}$, $A_1^{(1)}$, and $A_0^{(1)\ast}$ correspond to \eqref{eqn:mul_discrete}.
Let us consider the function $\widetilde{W}$ for the multiplicative surface of type $A_2^{(1)}$.
One of the discrete systems of type $A_2^{(1)}$ is
\begin{align}
 q\text{-}P(A_2)\quad
:&\left( {b_{1}\quad b_{2}\quad b_{3}\quad b_{4}\atop
b_{5}\quad b_{6}\quad b_{7}\quad b_{8}};f,g\right) \mapsto 
\left( {b_{1}\quad b_{2}\quad b_{3}\quad b_{4}\atop
qb_{5}\quad qb_{6}\quad qb_{7}\quad qb_{8}}
;\overline{f}, \overline{g}\right) ,\nonumber \\*
&\frac{(fg-1)(f\overline{g}-1)}{qb_{7}b_{8}}= 
\frac{(f-b_{1})(f-b_2)(f-b_3)(f-b_4)}{(f-b_{5})(f-b_{6})}, \nonumber\\*
&\label{eqn:qpa2}\frac{(f\overline{g}-1)(\overline{f}\overline{g}-1)}{qb_{5}b_{6}}=
\frac{(\overline{g}-1/b_1)(\overline{g}-1/b_{2})
(\overline{g}-1/b_3)(\overline{g}-1/b_4)}
{(\overline{g}-qb_{7})(\overline{g}-qb_{8})}, \\*
&\left( q=\frac{b_{5}b_{6}}{b_{1}b_{2}b_3b_4b_{7}b_{8}}\right) .
\nonumber
\end{align}
The symplectic form is $\omega =d\log G\wedge d\log F$, where $F=fg-1$, $G=g$.
Putting
\begin{align*}
 &\varphi(f,\overline{g})=\frac{1}{f}\left(1+qb_7b_8\frac{(f-b_1)(f-b_2)(f-b_3)(f-b_4)}
{(f\overline{g}-1)(f-b_5)(f-b_6)}\right),\\
 &\psi(f,\overline{g})=\frac{1}{\,\overline{g}\,}\left(1+qb_5b_6\frac{(\overline{g}-1/b_1)(\overline{g}-1/b_2)(\overline{g}-1/b_3)(\overline{g}-1/b_4)}
{(f\overline{g}-1)(\overline{g}-qb_7)(\overline{g}-qb_8)}\right),
\end{align*}
$\widetilde{W}(f,\overline{g})=W(f\varphi(f,\overline{g})-1,\overline{g})$ satisfies
\begin{align*}
 &\frac{\partial\widetilde{W}}{\partial f}
 =\frac{\partial {W}}{\partial F}\frac{\partial F}{\partial f}
 =\left(\varphi +f\frac{\partial\varphi}{\partial f}\right)\frac{\partial {W}}{\partial F},\quad
 \frac{\partial\widetilde{W}}{\partial\overline{g}}
 =\frac{\partial {W}}{\partial F}\frac{\partial F}{\partial\overline{g}}
 +\frac{\partial {W}}{\partial\overline{G}}
 =f\frac{\partial\varphi}{\partial\overline{g}}\frac{\partial {W}}{\partial F}
 +\frac{\partial {W}}{\partial\overline{G}}.
\end{align*}
Therefore, we have
\begin{align*}
 &\varphi =\exp\left(F\frac{\partial W}{\partial F}\right)
 =\exp\left(\frac{f\varphi -1}{\varphi +f\frac{\partial\varphi}{\partial f}}
 \frac{\partial\widetilde{W}}{\partial f}\right),\\
 &\psi =\frac{1+\exp\left(\overline{G}\frac{\partial W}{\partial\overline{G}}\right)}{\overline{G}}
 =\frac{1}{\,\,\overline{g}\,\,}\left(1+\exp\left(\overline{g}\frac{\partial\widetilde{W}}{\partial\overline{g}}
 -\frac{f\overline{g}\frac{\partial\varphi}{\partial\overline{g}}}{\varphi +f\frac{\partial\varphi}{\partial f}}
 \frac{\partial\widetilde{W}}{\partial f}\right)\right).
\end{align*}
 By integrating
\begin{align*}
 &\frac{\partial\widetilde{W}}{\partial f}=\frac{\varphi +f\frac{\partial\varphi}{\partial f}}{f\varphi -1}
 \log\varphi,\\
 &\frac{\partial\widetilde{W}}{\partial\overline{g}}=
 \frac{1}{\,\overline{g}\,}\log\left(\overline{g}\psi -1\right)
 +\frac{f\frac{\partial\varphi}{\partial\overline{g}}}{f\varphi -1}
 \log\varphi,
\end{align*}
we obtain $\widetilde{W}=\widetilde{W}_{\! A_2}$ as
\begin{align}
 \widetilde{W}_{\! A_2}=&
 \mathrm{Li}_2(f\overline{g})
 +\mathrm{Li}_2\left(\frac{(1-f/b_1)(1-f/b_2)(1-f/b_3)(1-f/b_4)}{(1-f\overline{g})(1-f/b_5)(1-f/b_6)}\right)
 +\log f\log (1-f\overline{g})\nonumber\\
 &+\sum_{k=1}^4\mathrm{Li}_2\left(1-\frac{f}{b_k}\right)-\sum_{l=5,6}\mathrm{Li}_2\left(1-\frac{f}{b_l}\right)
 +\sum_{k=1}^4\log b_k\log\left(1-\frac{f}{b_k}\right)-\sum_{l=5,6}\log b_l\log\left(1-\frac{f}{b_l}\right)
 \nonumber\\
 &-\sum_{k=1}^4\mathrm{Li}_2\left(b_k\overline{g}\right)
 +\sum_{l=7,8}\mathrm{Li}_2\left(\frac{\overline{g}}{qb_l}\right)
 +\pi\sqrt{-1}\log\overline{g}.
\end{align}
Using $\widetilde{W}$, the discrete system (\ref{eqn:qpa2}) is expressed as
\begin{align}
 &g=\exp\left(
 -\left(\sum_{k=1}^4\frac{1/b_k}{1-f/b_k}-\sum_{l=5,6}\frac{1/b_l}{1-f/b_l}
 -\frac{\overline{g}}{1-f\overline{g}}\right)^{-1}
 \frac{\partial\widetilde{W}}{\partial f}\right),\\
 &\overline{f}=\frac{1}{\,\overline{g}\,}
 \left(1+\exp\left(\overline{g}\frac{\partial\widetilde{W}}{\partial\overline{g}}
 -\frac{f\overline{g}}{1-f\overline{g}}
 \frac{\partial\widetilde{W}}{\partial f}\right)\right),
\end{align}
which is, however, probably not a satisfactory answer one hopes.



For the other cases, let us merely write coordinates $F$ and $G$.
In the cases of type $A_2^{(1)\ast}$, $A_1^{(1)\ast}$, and $A_0^{(1)\ast\ast}$, the coordinates
\begin{align}
  &A_2^{(1)\ast}:\ (F,G)=(f+g, g),\qquad
 A_1^{(1)\ast}:\!\ (F,G)=\left(1-\frac{2r}{f+g},g\right),\nonumber\\
 &A_0^{(1)\ast\ast}:\
 (F,G)=\left((f-g)^2-8r^2(f+g)+16r^4,f-g\right)
\end{align}
give the symplectic form $dG\wedge d\log F$.
For the surfaces of type $A_2^{(1)}$, $A_1^{(1)}$, and $A_0^{(1)\ast}$, the coordinates
\begin{align}
 &A_2^{(1)}:\ (F,G)=(fg-1,g),\qquad
 A_1^{(1)}:\ (F,G)=\left(1-\frac{r^2-1}{fg-1},g\right),\nonumber\\
 &A_0^{(1)\ast}:\
 (F,G)=
 \left((f+g)^2-\left(r^2+\frac{1}{r^2}\right)fg+r^2-\frac{1}{r^2},
 -\frac{f}{r}+rg\right)
\end{align}
give the symplectic form $d\log G\wedge d\log F$.

\section*{Acknowledgements}

This work is supported by Japan Society for the Promotion of Sciences (JSPS), through JSPS grants number
18K03323 and
18K13438.

\bibliographystyle{unsrt}

\bibliographystyle{plain}

\end{document}